%% file: hepex0001.tex
\documentstyle[epsfig,a4,12pt]{article}
\epsfverbosetrue

\def\nuebar{\rm{\bar{\nu_e}}}

\def\dm2{\rm{\Delta m^2}}
\def\munu{\mu_{\nu}}
\def\nurad{\rm{ < r^2 > }}
\def\s2tw{\rm{ sin ^2 \theta _W }}
\def\am241{\rm{ ^{241} Am }}
\def\u238{\rm{ ^{238} U }}
\def\th232{\rm{ ^{232} Th }}
\def\k40{\rm{ ^{40} K }}

\def\cs133{\rm{ ^{133} Cs }}
\def\i127{\rm{ ^{127} I }}
\def\li7{\rm{ ^{7} Li }}
\def\b2o3{\rm{B_2 O_3}}

\begin{document}

\hfill AS-TEXONO/00-01 \\
%%\hspace*{1cm} \hfill \today

\begin{center}
\large
\bf{
A CsI(Tl) Scintillating Crystal Detector for the Studies of \\
Low Energy Neutrino Interactions \\
}
\vspace*{0.5cm}
\normalsize
H.B. Li$^{a,b}$,
Y. Liu$^c$,
C.C. Chang$^d$,
C.Y. Chang$^d$,
J.H. Chao$^e$,
C.P. Chen$^a$,
T.Y. Chen$^f$,\\
M. He$^g$,
L. Hou$^h$,
G.C. Kiang$^a$,
W.P. Lai$^a$,
S.C. Lee$^a$, 
J. Li$^c$,          
J.G. Lu$^c$,
Z.P. Mao$^c$,\\
H.Y. Sheng$^{a,c}$,
R.F. Su$^i$,
P.K. Teng$^a$, 
C.W. Wang$^a$,
S.C. Wang$^a$,
H.T. Wong$^{a,}$\footnote{Corresponding~author:
Email:~htwong@phys.sinica.edu.tw;
Tel:+886-2-2789-9682;
FAX:+886-2-2788-9828.},\\
T.R. Yeh$^j$,
Z.Y. Zhang$^k$,
D.X. Zhao$^{a,c}$,
S.Q. Zhao$^c$,
Z.Y. Zhou$^h$,
B.A. Zhuang$^{a,c}$ \\[2ex]
{\large
The TEXONO\footnote{Taiwan EXperiment On NeutrinO} Collaboration 
}
\end{center}

\normalsize

\begin{flushleft}
$^a$ Institute of Physics, Academia Sinica, Taipei, Taiwan.\\
$^b$ Department of Physics, National Taiwan University, Taipei, Taiwan.\\
$^c$ Institute of High Energy Physics, Beijing, China.\\
$^d$ Department of Physics, University of Maryland, College Park, U.S.A.\\
$^e$ Nuclear Science and Technology Development
Center, National Tsing Hua University, \\
\hspace*{1cm} Hsinchu, Taiwan.\\
$^f$ Department of Physics, Nanjing University, Nanjing, China.\\
$^g$ Department of Physics, Shandong University, Jinan, China.\\
$^h$ Department of Nuclear Physics, 
Institute of Atomic Energy, Beijing, China.\\
$^i$ Nuclear Engineering Division, Nuclear Power Plant II, Kuosheng, Taiwan.\\
$^j$ Engineering Division, Institute of Nuclear Energy Research,
Lungtan, Taiwan.\\
$^k$ Department of Electronics,
Institute of Radiation Protection, Taiyuan, China.\\
\end{flushleft}

\vfill

\pagebreak

\begin{center}
{\bf
Abstract
}
\end{center}

Scintillating crystal detector may offer
some potential advantages in the 
low-energy,
low-background experiments.
A 500~kg CsI(Tl) detector 
to be placed near the core
of Nuclear Power Station II in Taiwan
is being constructed for the 
studies of electron-neutrino 
scatterings and other keV$-$MeV range
neutrino interactions.
The motivations of this detector 
approach, the physics to be
addressed, the basic experimental design, and 
the characteristic performance of prototype modules
are described. The expected background channels
and their experimental handles are discussed.

\vspace*{0.1cm}

\begin{flushleft}
\end{flushleft}

\newpage

\setlength{\headsep}{-1cm}

\section{Introduction}

Scintillating crystal detectors~\cite{crystal} have been
widely used as electromagnetic calorimeters in
high energy physics~\cite{emcalo}, as well as in medical and
security imaging
and in the oil-extraction industry.
They offer many potentials merits~\cite{prospects} 
for low-energy (keV-MeV range) low background experiments.

A CsI(Tl) scintillating crystal detector
is being constructed
to be placed near a reactor core 
to study low energy neutrino interactions~\cite{pilot}.
In subsequent sections, we discuss
the motivations for this choice of
detector technology, 
the physics topics to be addressed,
the basic experimental design
and the prototype performance parameters.
The various background processes and the
experimental means to identify and suppress
them are considered.
Various extensions based
on this detector technique are
summarized at the end.

\section{Physics and Detector Motivations}
\label{sect::det}

High energy (GeV) neutrino beams from accelerators
have been very productive in the 
investigations of electroweak,
QCD and structure function
physics~\cite{nuint} and have blossomed into
matured programs at CERN and Fermilab. 
However, the use of low energy (MeV) neutrino
as a probe to study particle and nuclear physics
has not been well explored. 

Nuclear power reactors  are abundant source of 
electron anti-neutrinos ($\nuebar$) at the MeV range. 
Previous experiments with reactor neutrinos
primarily focused on the 
interactions
\begin{displaymath}
\rm{ \bar{\nu_e}~ + ~ p ~  \rightarrow ~ e^+ ~ +  ~ n} 
\end{displaymath}
to look for neutrino oscillations~\cite{refrnuexpt}.
The choice of this interaction channel is
due to its relatively large cross-sections,
the very distinct experimental signatures (prompt
e$^+$ followed by a delayed neutron capture),
and the readily available detector
choice of liquid scintillator providing the proton target.
Using this interaction, the reactor neutrino spectrum
has been measured to a precision of $\sim$2\% in
the 3~MeV to 7~MeV range from
the Bugey experiment~\cite{bugeyspect}.

The only other processes measured at the MeV range
for $\nuebar$ are $\nuebar$-electron~\cite{reinesnue,kurtchatovnue,rovnonue} 
and $\nuebar$-deuteron~\cite{nudexpt}
interactions, and their accuracies are at
the 30-50\% and 10-20\% range, respectively.
There are motivations to improve on these
measurements, to investigate
complementary detection techniques,
and to study the other unexplored channels.

Gamma-ray spectroscopy has been a
standard technique in nuclear sciences
(that is, in the investigations of physics
at the MeV range), with the use of scintillating
crystals or solid-state detectors. 
Gamma-lines of characteristic
energies give unambiguous information on the
presence and transitions of whichever specific
isotopes, allowing a unique interpretation of both the 
signal and background processes. 

Several intrinsic properties of 
crystal scintillators make them
attractive candidates for low-background
experiments~\cite{prospects} in neutrino
and astro-particle physics.
The experimental difficulties of building a
high-quality gamma detector for MeV neutrino
physics have been the large target mass required.
However, in recent years, big electro-magnetic
calorimeter systems~\cite{emcalo}
(such as the mass of 40 tons of CsI(Tl) 
crystals, in the case
for the current B-factories experiments~\cite{bfactories})
have been built for high energy physics experiments,
using CsI(Tl)
crystals~\cite{csichar} with 
silicon PIN photo-diodes readout~\cite{sipd}.
In addition, NaI(Tl) detectors with the mass range of 100~kg
have been used in Dark Matter WIMP 
searches~\cite{dmnai}, producing
some of the most sensitive results.

The CsI-crystal production technology is by now well matured
and the cost has been reduced enormously due to the
large demands. It becomes realistic and affordable
to build a CsI detector in the range of 1-ton
in target mass
for a reactor neutrino experiment.
The detector is technically
much simpler to build and to operate than, for instance,
gas chambers and liquid scintillators.
The detector mass
can be readily scaled up to tens of tons 
if the first experiment would
yield interesting results or lead to
other potential applications.
Other scintillating crystal detectors 
can be easily customized
in the various potential applications.

The properties of
CsI(Tl) crystals, together with those
of a few common scintillators, are listed in
Table~\ref{scintab}. The CsI(Tl) crystal offers
certain advantages over the other possibilities.
It has relatively high light yield
and high photon absorption
(or short radiation length). It is mechanically
stable and easy to machine, and is only
weakly hygroscopic. 
There is no need for a hermetic container to seal
the detector from the ambient humidity (as
required, for instance,
by NaI(Tl)). This minimizes radioactive
background as well as energy loss in the
passive elements which will degrade
energy resolution.
In particular, CsI(Tl) provides
strong attenuation for $\gamma$'s of energy less 
than 500~keV. 
As a result, it is possible to realize
a compact detector design with 
minimal passive materials within
the fiducial volume, 
and with an efficient external
shielding configuration.

\section{Neutrino Interactions on CsI(Tl)}
\label{sect::phys}

\subsection{Neutrino-Electron Scattering}

The cross section
for the process
\begin{displaymath}
\rm{
\bar{\nu_e} ~ + ~ e^- ~ \rightarrow ~ \bar{\nu_e} ~ + ~ e^- 
}
\end{displaymath}
gives information on the electro-weak parameters 
($\rm{ g_V , ~ g_A , ~ and ~ sin ^2 \theta_W }$), 
and are sensitive to 
small neutrino magnetic moments ($\munu$) 
and the mean square charge 
radius ($\nurad$)~\cite{kyuldjiev,vogelengel}.
Scatterings of the 
$\rm{( \nu_e ~ e )}$~\cite{nueexptlampf}
and
$\rm{( \nuebar ~ e )}$ 
are two of the most
realistic systems
where the interference
effects between Z [neutral currents (NC)] and 
W [charged currents (CC)] exchanges 
can be studied~\cite{kayser}.
Many of the existing
and proposed solar neutrino detectors 
(Super-Kamiokande, Borexino, HELLAZ, HERON ...)~\cite{snuexpt}
make use of the $\rm{\nu_e}$-e interactions
as their detection mechanisms.
The fact the $\rm{\nu_e}$-e scattering can proceed
via both W and Z exchanges,
while $\rm{\nu_{\mu}}$-e and $\rm{\nu_{\tau}}$-e
are purely neutral current processes,  is the
physics-basis of Resonance Neutrino
Oscillation in Matter (the ``MSW'' Effect)~\cite{msw}.

In an experiment, what can be measured is the recoil
energy of the electron (T). The differential cross
section can be expressed as :
\begin{eqnarray*}
\frac{ d \sigma }{ dT } ( \nu ~ e ) & =  &
\frac{ G_F^2 m_e }{ 2 \pi } 
[ ( g_V \pm x \pm g_A )^2 + ( g_V \pm x \mp g_A )^2  
[ 1 - \frac{T}{E_{\nu}} ]^2
+ ( g_A^2 - ( g_V \pm x )^2  ) \frac{ m_e T }{E_{\nu}^2}  ] \\
 &  &  + \frac{ \pi \alpha _{em} ^2 \munu ^2 }{ m_e^2 }
[ \frac{ 1 - T/E_{\nu} }{T} ]
\end{eqnarray*}
where 
$ g_V = 2 ~ \s2tw - \frac{1}{2}$ and $ g_A =  - \frac{1}{2}$
in the Standard Model,
while the upper (lower) signs refer to 
$ \nu_{\mu} ~ e$ and $\nu_{\tau}~ e$ 
($ \bar{\nu_{\mu}} ~ e$ and $\bar{\nu_{\tau}}~ e$) scatterings
where only NC are involved,
and
\[
x = \frac{2 M_W^2}{3} \nurad \s2tw ~~ .
\]
For $\nu_e ~ e$ scattering, both NC and CC
and their interference terms contribute, so that the cross sections
can be evaluated by replacing 
$g_V \rightarrow g_V + 1$ and 
$g_A \rightarrow g_A + 1$.
The $\munu$ term have a 
$\rm{\frac{1}{T}}$ dependence. 
Accordingly, experimental searches for
the neutrino magnetic moment should 
focus on the reduction of the threshold
(usually background-limited) for 
the recoil electron energy.

\begin{figure}
\centerline{
\epsfig{file=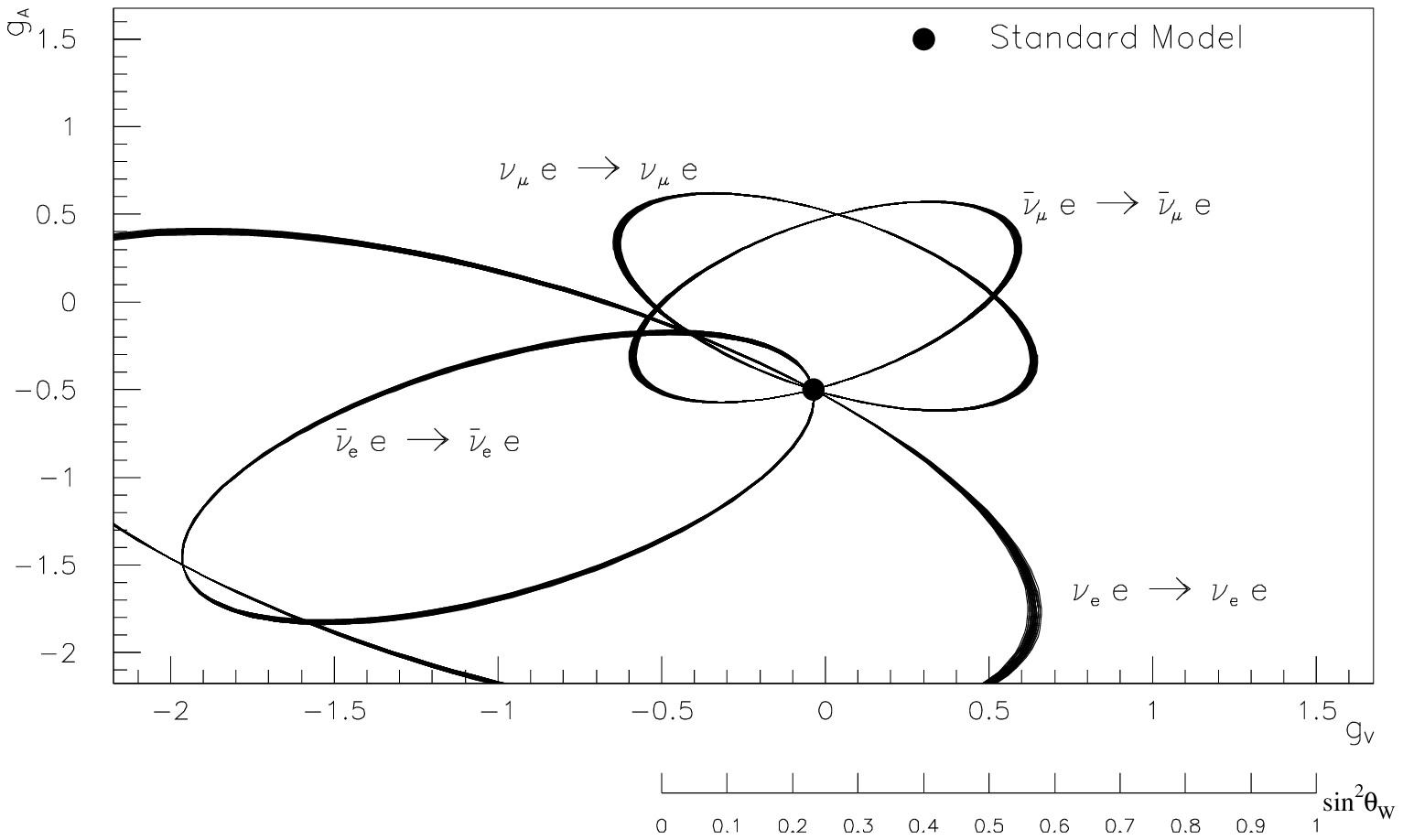,width=16cm}
}
\caption{
Sensitivities of typical ($\nu_{\mu}$ e),
($\bar{\nu_{\mu}}$ e), ($\rm{\nu_e}$ e) 
and ($\nuebar$ e) cross-section measurements to the
different regions in the $\rm{g_A}$-$\rm{g_V}$ parameter
space, showing
their complementarity. The Standard Model values are denoted
by the black dot.
}
\label{gvvsga}
\end{figure}

\begin{figure}
{\bf (a)}
\centerline{
\epsfig{file=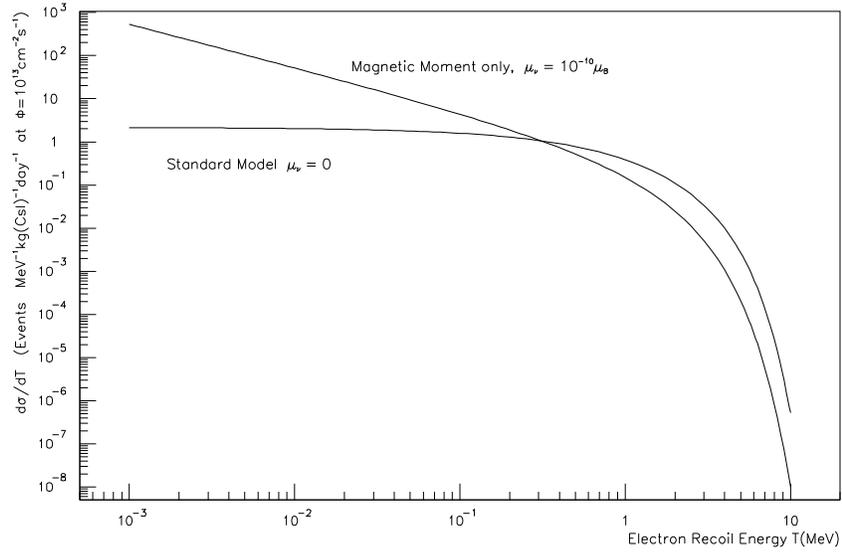,width=13cm}
}
{\bf (b)}
\centerline{
\epsfig{file=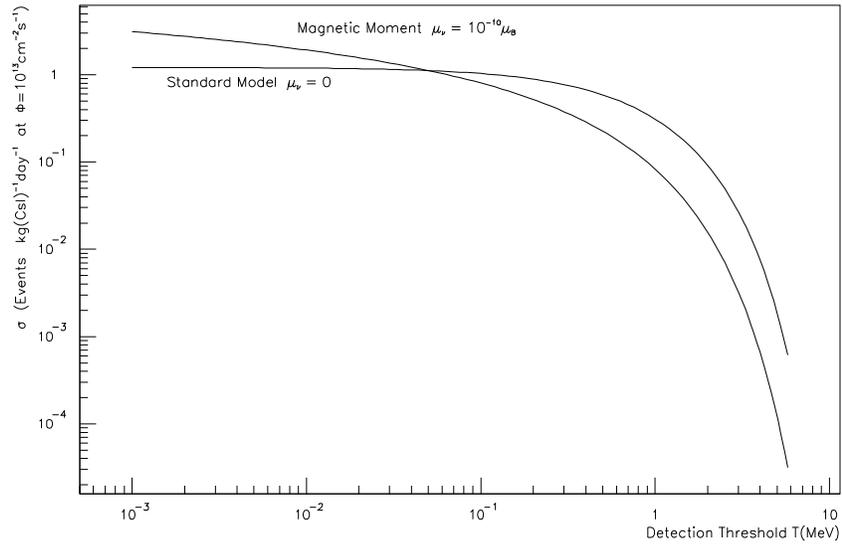,width=13cm}
}
\caption{
(a)
Differential cross section showing the
electron recoil energy spectrum in
$\nuebar$-e scatterings, and 
(b)
expected event rates as a function
of the detection threshold of the 
recoil electrons, 
at a reactor
neutrinos flux of $\rm{10^{13}~cm^{-2} s^{-1}}$, 
for the Standard Model processes and
for the contribution due to a neutrino
magnetic moment of 10$^{-10}$ Bohr magneton.
}
\label{nuerecoil}
\end{figure}

The $\rm{g_A ~ Vs. ~ g_V}$ parameter space where
$\rm{( \nuebar ~ e )}$ scatterings are sensitive to
is depicted in Figure~\ref{gvvsga}. 
The complementarity
with $\rm{ ( \nu_{\mu} ~ e , ~ \bar{\nu_{\mu}}  ~ e , \nu_{e} ~ e ) }$ 
can be readily seen. 
The expected recoil energy spectrum~\cite{vogelengel}
is displayed in Figure~\ref{nuerecoil}a, showing
standard model expectations and the case with
an anomalous neutrino magnetic moment of 
$\rm{10^{-10}~\mu_B}$.
The present published limit~\cite{rovnonue} is
$\rm{1.9 \times 10^{-10}~\mu_B}$.
The number of events in the two cases
as a function of measurement threshold 
is depicted in Figure~\ref{nuerecoil}b.
It can be seen that the
rate is at the range
of O(1) events per kg of CsI(Tl) per day 
[$\equiv$  ``pkd'']\footnote{For simplicity,
we denote {\it ``events per kg of CsI(Tl) per day''} by 
{\bf pkd} 
in this article.}
at about 100~keV threshold at a reactor
neutrino flux of 
$\rm{10^{13} ~ cm^{-2} s^{-1}}$,
which poses formidable experimental
challenge in terms of background control.

Therefore, investigations of $\rm{( \nuebar ~ e )}$
cross-sections with reactor neutrinos
allow one to study electro-weak physics 
at the MeV range, to probe charged and neutral
currents interference,
and to look for an anomalous neutrino magnetic moment.  
The present experimental situations are discussed
in Ref.~\cite{numunim} and summarized in 
Table~\ref{tabnueexpt}. 
In particular, 
a re-analysis~\cite{vogelengel} 
of the Savannah River results~\cite{reinesnue},
based on an improved reactor neutrino spectrum
and the Standard Model $\s2tw$ value,
suggested that the measured ($\nuebar$ e)
cross-sections at 1.5-3.0~MeV and 3.0-4.5~MeV
are 1.35$\pm$0.4 and 2.0$\pm$0.5  times, respectively,
larger than the expected values, 
and can be interpreted
to be consistent with a $\rm{\mu_{\nu}}$ at the range
of (2-4)X$\rm{10^{-10}~\mu_B}$.
Various astrophysics considerations from the
time duration of the supernova SN1987A burst~\cite{sn1987nue},
stellar cooling~\cite{starcoolnue} and
Big Bang Nucleosynthesis~\cite{bbnnue}
provide more stringent bounds on
$\munu$ to the
$\rm{10^{-11} - 10^{13} ~ \mu_B }$ level,
but these are model-dependent.
An anomalous neutrino magnetic moment
of range $\rm{\mu_{\nu} \sim 10^{-10}~\mu_B }$
has been considered 
as a solution to the Solar Neutrino Puzzle~\cite{snunue}.
There are motivations to improve the 
cross-section measurements and magnetic
moment sensitivities further with
laboratory experiments. 
Several other projects are underway~\cite{numunim,refnueexpt}.

A 500~kg CsI(Tl) crystal calorimeter (fiducial
mass 200-300~kg) will have more
target electrons than previous 
experiments~\cite{reinesnue,kurtchatovnue,rovnonue} and 
current projects~\cite{numunim,refnueexpt},
as shown in Table~\ref{tabnueexpt}.
The signature for  $\rm{( \nuebar ~ e )}$
will be a single hit 
out of the several hundred channels.
As discussed in Section~\ref{sect::bkg},
the crystal scintillator approach may provide
low detection threshold, high photon
attenuation, and powerful
diagnostic capabilities for
background understanding. 
All these features 
can potentially improve the
sensitivities for
both cross-section measurements and
magnetic moments studies.

\subsection{Neutrino Interactions on $\cs133$ and $\i127$}

Neutral current excitation (NCEX) on nuclei by neutrinos 
\begin{displaymath}
\rm{
\bar{\nu_e} ~  +   ~ (A,Z) ~
\rightarrow ~ \bar{\nu_e} ~ + ~ (A,Z)^* ~
}
\end{displaymath}
has been observed only in the case of $^{12}$C~\cite{karmennuex}
with intermediate energy (O(10~MeV)) 
neutrinos. Excitations with lower energies
using reactor neutrinos have been studied theoretically~\cite{nuex}
but not observed.

Crystal scintillators, having good $\gamma$ resolution
and capture efficiency, are suitable to study these processes.
The experimental signatures will be  an excess
of events at the characteristic gamma-energies
correlated to the Reactor-ON period.
Using CsI(Tl) as
active target nuclei, the candidate $\gamma$-lines 
with M1 transitions
include 81 and 160~keV for $\cs133$, and
58, 202 and 418~keV for $\i127$. 
The use of NCEX as the detection mechanisms 
for solar neutrinos~\cite{b11nuex,lii} and
Dark Matter-WIMPs~\cite{wimpncex} have been
discussed. Competitive limits have been set with
the WIMP-searches based on the
NCEX channel with $\i127$~\cite{wimpi127}
and $^{129}$Xe~\cite{wimpxe129}.

There are no theoretical predictions
for these transitions for $\cs133$ and $\i127$. 
One may expect a similar range as
the 480~keV case for $^7$Li~\cite{nuex},
which would be O(0.01-0.1)~pkd at a reactor neutrino
flux of $\rm{10^{13}~cm^{-2} s^{-1}}$.
In addition, there are 
theoretical work~\cite{nuexaxial}
suggesting that
the NCEX cross-sections on $^{10}$B and $^{11}$B
are sensitive to the
axial isoscalar component of NC interactions
and the strange quark content of the nucleon.
The studies of neutrino-induced interactions on
nuclei is one of the principal program of the
ORLaND proposal~\cite{orland} based on
intermediate energy neutrinos
from spallation neutron source.

For completeness, we mention that $\nuebar$ can also interact
with $\cs133$ and $\i127$ via the  charged-current (CC) channels.
There are two modes:
(I) Inverse beta decay
\begin{displaymath}
\rm{
\bar{\nu_e} ~  +   ~ (A,Z) ~
\rightarrow ~ e^+ ~ + ~ (A,Z-1)^* ~
}
\end{displaymath}
have the distinct signatures of 2 back-to-back 
511~keV $\gamma$s from
the positron annihilation, plus the characteristic $\gamma$-lines
from the daughter nuclei, and the positron itself; 
(II) Resonant orbital electron
capture
\begin{displaymath}
\rm{
\bar{\nu_e} ~  + ~ e^- ~ + ~ (A,Z) ~
\rightarrow ~ ~ (A,Z-1)^* ~
}
\end{displaymath}
takes place only at a narrow range of
neutrino energy equal to the Q-value of the transition.
The signatures are the characteristic gamma lines for the
excited daughter nuclei. Calculation does not exist
but the event rates  are
expected to be further suppressed since they
involve the conversion of a proton to a neutron
in neutron-rich nuclei.
The $\nuebar$N-CC interactions have been considered to
detect the low-energy terrestrial neutrinos due to the radioactivity
at the Earth's lithosphere~\cite{earthnu}.

\section{Experimental Design}

Since mid-1997, a Collaboration has been 
built up~\cite{taiwan} to pursue an experimental
program discussed in Section~\ref{sect::phys}
$-$ the studies of low energy neutrino interactions
using reactor neutrinos as source
with CsI(Tl) crystal as detector~\cite{pilot}. 

\begin{figure}[p]
\centerline{
\epsfig{file=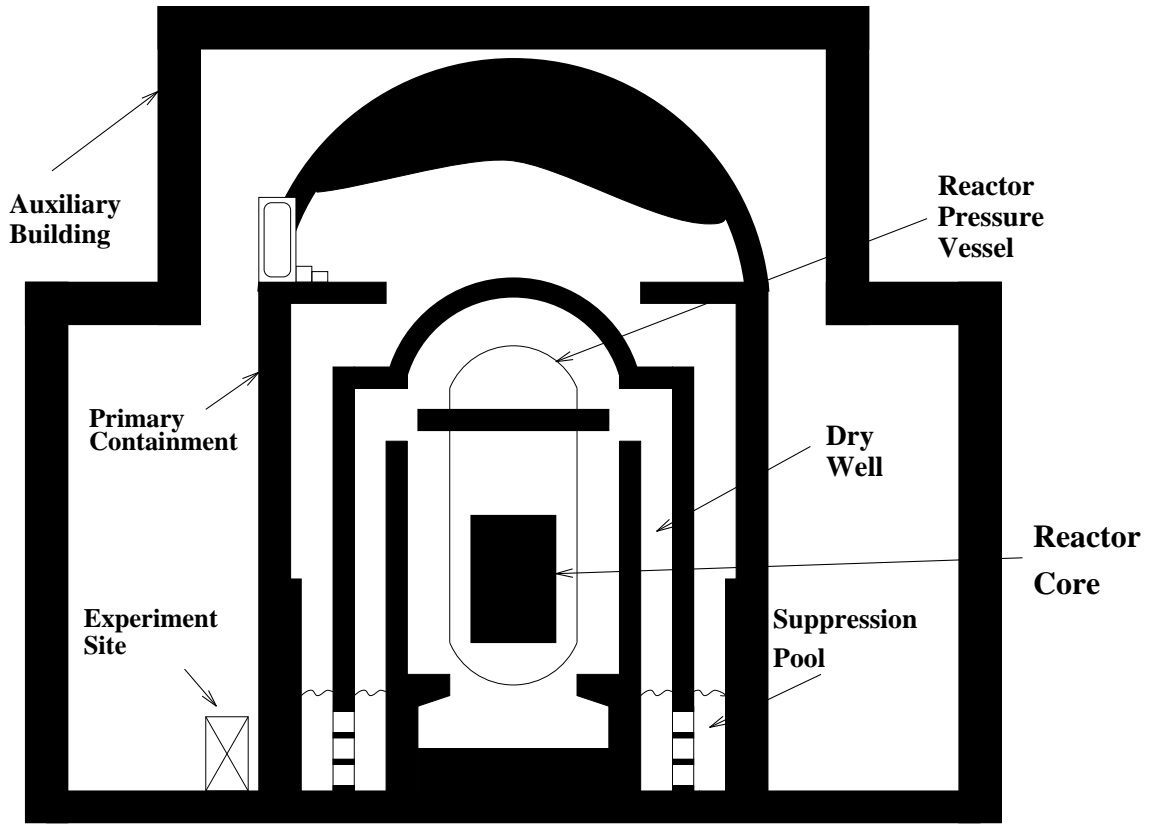,width=12cm,angle=270}
}
\caption{
Schematic side view, not drawn to scale,
of the NP2 Reactor Building,
indicating the experimental site.
The reactor core-detector distance is about
28~m.
}
\label{fplanside}
\end{figure}

The experiment will be performed at the Nuclear
Power Station II at Kuo-sheng at the northern
shore of Taiwan. The experimental location is about
28~m from one of the reactor cores, and 102~m from
the other one. Each of the cores is a boiling water
reactor with 2.9~GW thermal power output, giving
a total flux of about $\rm{5.6 \times 10^{12} ~ cm^{-2} s^{-1}}$
at the detector site. The site is at the lowest level of
the reactor building, with about 25~mwe of overburden,
as depicted schematically in Figure~\ref{fplanside}.

To fully exploit the advantageous features
of the scintillating crystal approach~\cite{prospects}
in low-energy low-background experiments,
the experimental configuration
should enable the definition of a fiducial
volume with a surrounding active 4$\pi$-veto,
and minimal passive materials.

\begin{figure}
\centerline{
\epsfig{file=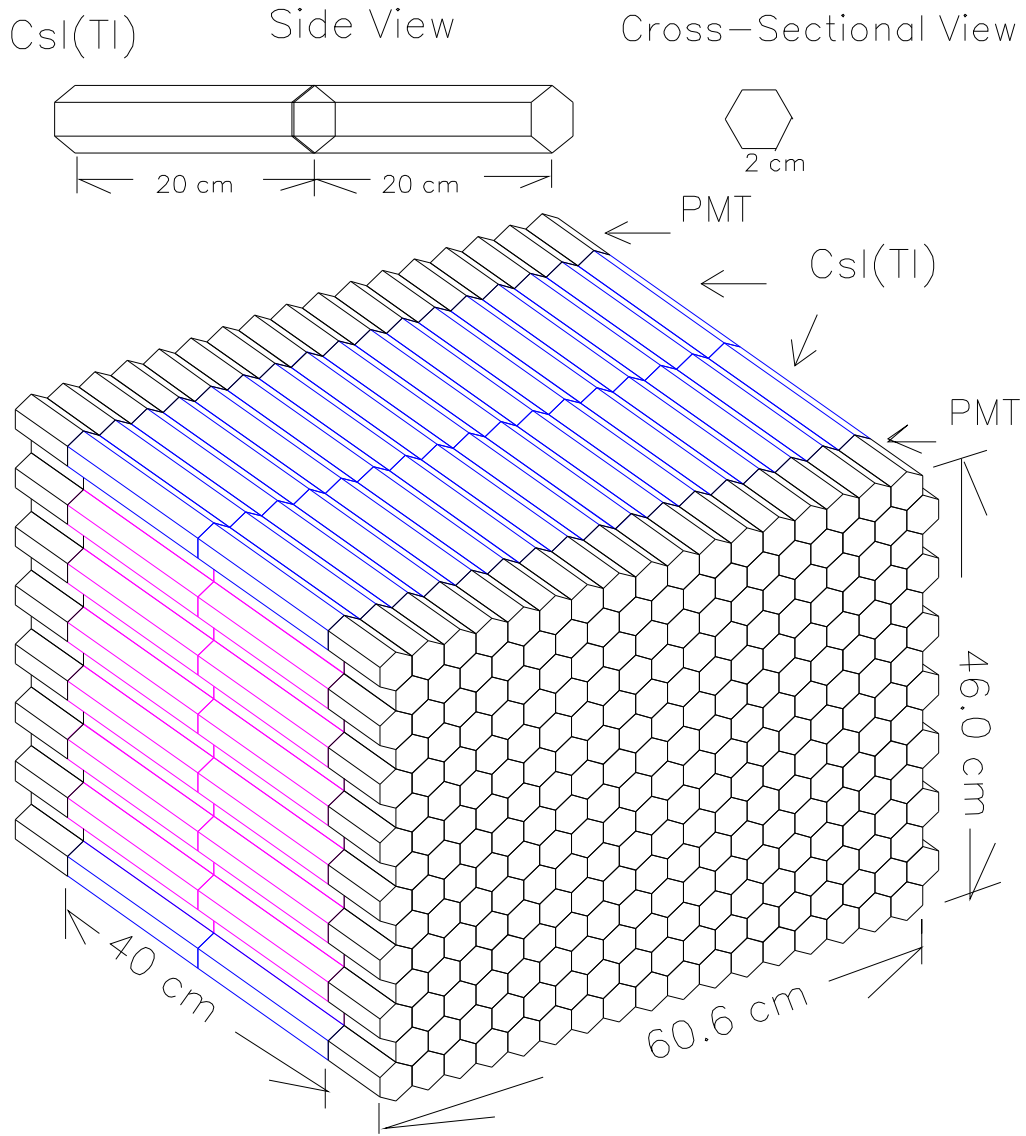,width=15cm}
}
\caption{
Schematic drawings of the CsI(Tl)
target configuration, showing a 2(Width)X
17(Depth)X15(Height) matrix.
Individual crystal module is 20~cm long with
a hexagonal cross-section of 2~cm edge. Readout
is performed by photo-multipliers 
at both ends.
}
\label{csitarget}
\end{figure}

The schematic design of the experiment is 
shown in Figure~\ref{csitarget}.
The detector will consist of about 480~kg of CsI(Tl) 
crystals\footnote{Manufacturer: Unique Crystals, Beijing}, 
arranged in a $17 \times 15$ matrix. 
One CsI(Tl) crystal unit
consists of a hexagonal-shaped cross-section with 2~cm
side and a length 20~cm, giving a mass of 0.94~kg.
Two such units are glued optically
at one end to form a module. 
The light output
are read out 
at both ends  by
29~mm diameter
photo-multipliers (PMTs) with 
low-activity glass window~\footnote{Hamamatsu CR110 customized},
which provide about 50\% of end-surfaces coverage.
The design of the PMT base is optimized for
high gain (low threshold) and good linearity over
a large dynamic range.
The modular design enables the detector
to be constructed in stages.

Individual crystals are wrapped with 
4 layers of  70~$\mu$m thick teflon sheets 
to provide diffused reflection for optimal light collection.
The sum of the two PMT signals 
($\rm{Q_{tot}=Q_1+Q_2}$) gives the
energy of the event, while the difference will provide
a measurement of a longitudinal position.
Outer layers as well as the last few cm near
the readout surfaces will be used as active veto.
The exact definitions of fiducial and veto volume can
be fine-tuned based on the actual background, and
can differ with different energy ranges.

\begin{figure}[p]
\centerline{
\epsfig{file=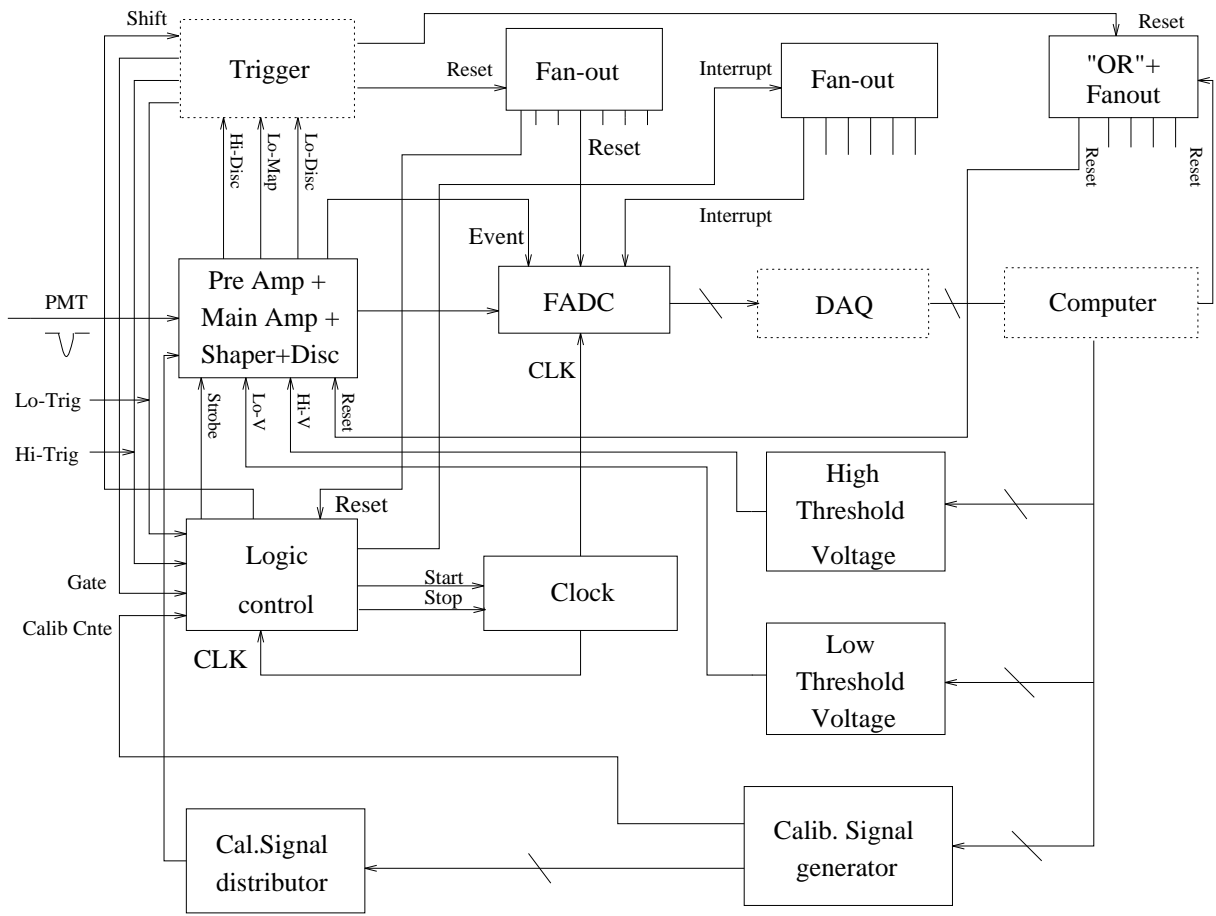,width=12cm,angle=270}
}
\caption{
Schematic layout of the electronics
system.
}
\label{electronics}
\end{figure}

\begin{figure}[p]
\centerline{
\epsfig{file=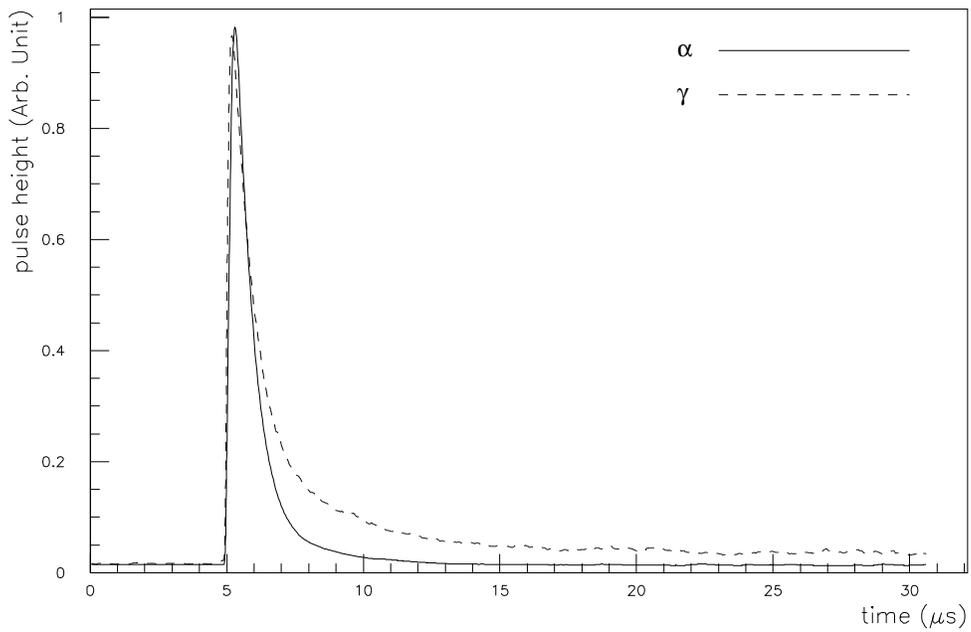,width=15cm}
}
\caption{
The average pulse shape
of events due to $\gamma$-rays and $\alpha$-particles
as recorded by the
FADC module. Their different decay times provide
pulse shape discrimination capabilities.
}
\label{pulse}
\end{figure}

The schematics of the electronics system
is depicted in Figure~\ref{electronics}.
The PMT signals are fed to 
amplifiers and shapers, and
are finally digitized by 8-bit
Flash-Analog-Digital-Convertor
(FADC) modules running at a clock rate of 20~MHz. 
The shaping is optimized for the $\mu$s time-scale
rise and fall times, such that noise-spikes from
single photo-electron are smeared out and suppressed.
Typical scintillation pulses due to
$\gamma$ and $\alpha$ events as
measured by the system are displayed in Figure~\ref{pulse}.
A precision pulse generator provides means to calibrate
and monitor the performance and stability
of the electronics system.

The
trigger conditions include: (a)
having any one or more channels above
a pre-set ``high threshold'' (typically 50-100~keV equivalent),
and (b) not having a cosmic veto signal within a previous
time-bin of typically 100~$\mu$s.
Once these conditions are fulfilled, all channels with 
signals above a ``low threshold'' (typically 10-30~keV equivalent)
will be read out.
The logic control circuit
enables complete acquisition of delayed signatures
up to several ms, to record cascade events in decay
series like $\u238$ and $\th232$.

\begin{figure}
\centerline{
\epsfig{file=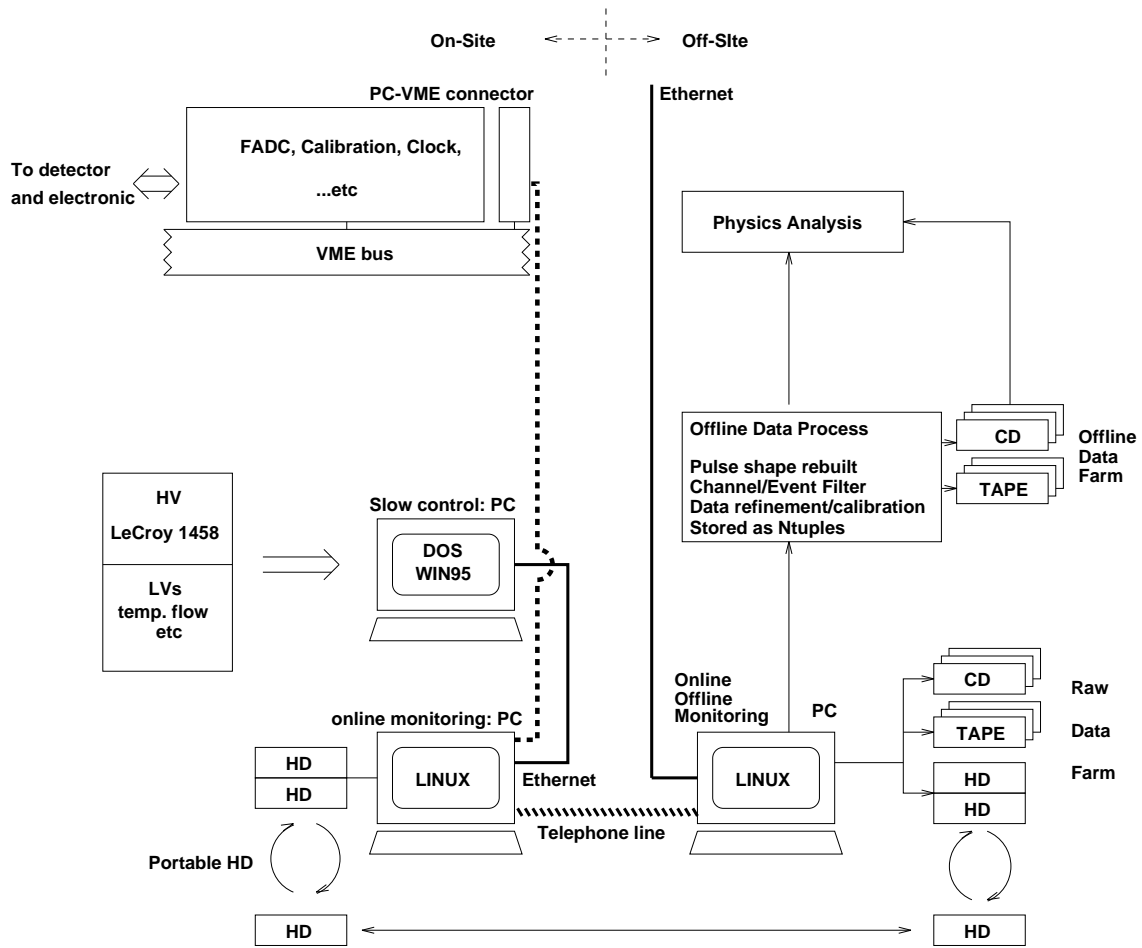,width=15cm}
}
\caption{
Schematic layout of the data acquisition
and on-line monitoring systems, and its interfacing
with the off-line software packages. A telephone line
provides connection from home-base laboratories
to the experimental site in the reactor building.
}
\label{software}
\end{figure}

The FADC, the 
trigger units, logic control and calibration
modules, are read out and controlled
by a VME-based data acquisition system,
connected by a PCI-bus to a PC running
with the LINUX operating system.
The on-line and off-line software architecture,
together with their inter-connections, 
are shown schematically in Figure~\ref{software}.
The on-site data taking conditions can be
remotely monitored from the home-base
laboratories via telephone line. (Internet access
to the Nuclear Power Plant is not allowed.)
Data will be saved in hard disks on-site and 
replaced at the once-per-week interval. 
They are duplicated and stored in magnetic tapes
and CDs for subsequent off-line analysis.
The detailed design and performance of the electronics,
data acquisition and control systems will
be the subject of a forthcoming article.

\begin{figure}
\centerline{
\epsfig{file=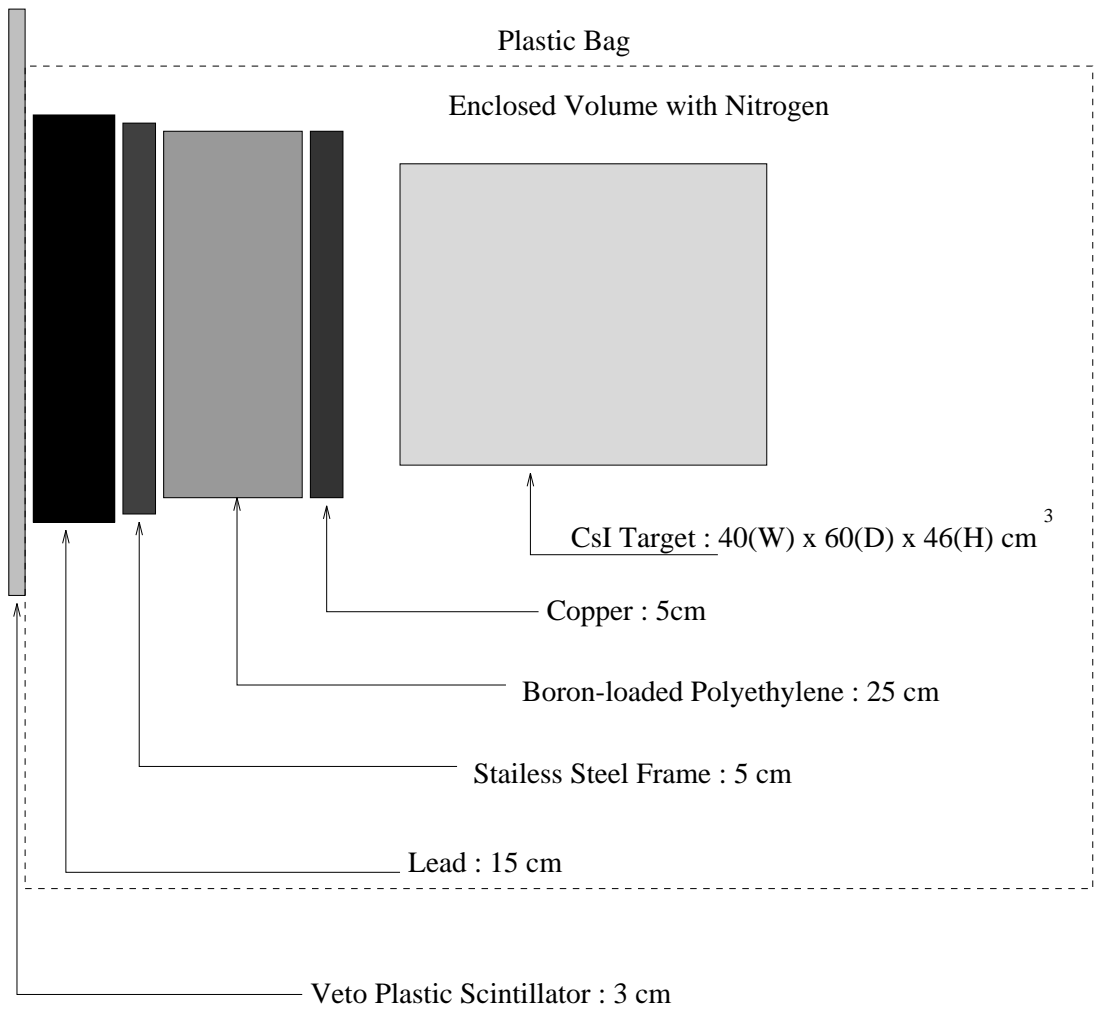,width=15cm,angle=270}
}
\caption{
Schematic layout of the target and shielding
configuration.
The coverage is 4$\pi$ but only one face
is shown.
}
\label{shielding}
\end{figure}

The compact CsI(Tl) detector enables an 
efficient shielding to be built.
The schematics of the shielding configuration
is depicted in Figure~\ref{shielding}.
Cosmic-rays and their
related events will be vetoed by
an outermost layer of plastic scintillators.
The typical veto gate-time will be $\sim$100~$\mu$s
to allow for delayed signatures due to neutron
interactions.
Ambient radioactivity is suppressed by
15~cm of lead and 5~cm of steel. The steel
layer also provides the mechanical structures to the 
system. Neutrons, mostly cosmic-induced 
in the lead and steel, are slowed down
and then absorbed by 25~cm of boron-loaded
polyethylene. The inner 5~cm of 
oxygen-free-high-conductivity (OFHC) copper
serves to suppress residual radioactivity from
the shielding materials.
The copper layers can be dismounted and
replaced by more polyethylene, allowing
flexibilities to optimize the shielding conditions
with respect to different physics focus.
The CsI(Tl) target will be placed inside 
a electrically-shielded
and air-tight box made of copper sheet.
The entire inner target space will be covered by
a plastic bag flushed with dry nitrogen
to prevent the radioactive radon gas from
diffusing into the target region.

To enable detector access  and maintenance,
the entire shielding assembly consists of three
parts: (1) a fixed shielding house, (2) a
movable trolley on which the target detector sits,
and (3) the front door which can
be moved by wheels. 
All the access pipes and  cable trays
are bent so
that there are no direct line-of-sight between
the inner target and the external background.
Ports are provided to allow insertion
of radiation sources for regular monitoring 
and calibration. These ports are blocked by 
copper plugs during normal data taking.

\section{Performance of Prototype Modules}

\begin{figure}
\centerline{
\epsfig{file=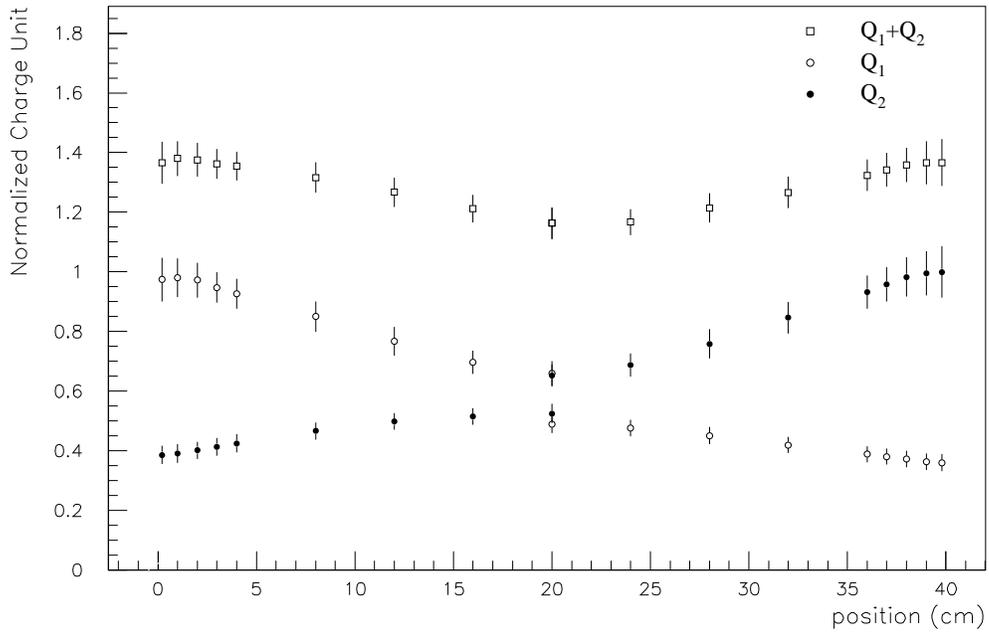,width=15cm}
}
\caption{
The measured variations of
Q$_1$, Q$_2$ and $\rm{Q_{tot}= Q_1 + Q_2}$ along
the longitudinal position of the crystal module.
The charge unit is normalized to unity for both
Q$_1$ and Q$_2$ at their respective ends.
The error bars denote the width  of the
photo-peaks due to a $^{137}$Cs source.
}
\label{qvsz}
\end{figure}

Extensive measurements on the crystal prototype
modules have been performed.  
The response is depicted
in Figure~\ref{qvsz}, showing the variation
of collected light for $\rm{Q_1}$, $\rm{Q_2}$ 
and $\rm{Q_{tot}}$ as a function of position
within one crystal module.
The charge unit is normalized to unity at the
$^{137}$Cs photo-peak (660~keV) for both
Q$_1$ and Q$_2$ at their respective ends, while
the error bars denote the FWHM width at that energy.
The discontinuity at L=20~cm
is due to the optical mis-match between the
glue (n=1.5) and the CsI(Tl) crystal (n=1.8).
It can be seen that there is a dependence
of $\rm{Q_{tot}}$ with position at the 10-20\% level.
A FWHM energy resolution of 10\% is achieved  at
660~keV, and its variation with energy follows 
the $\rm{E^{- \frac{1}{2}}}$ relation.
The detection threshold (where signals
are measured at both PMTs) is $<$20~keV.
A good linearity of the PMT response
is achieved for energies
from 20~keV to 20~MeV.

\begin{figure}
\centerline{
\epsfig{file=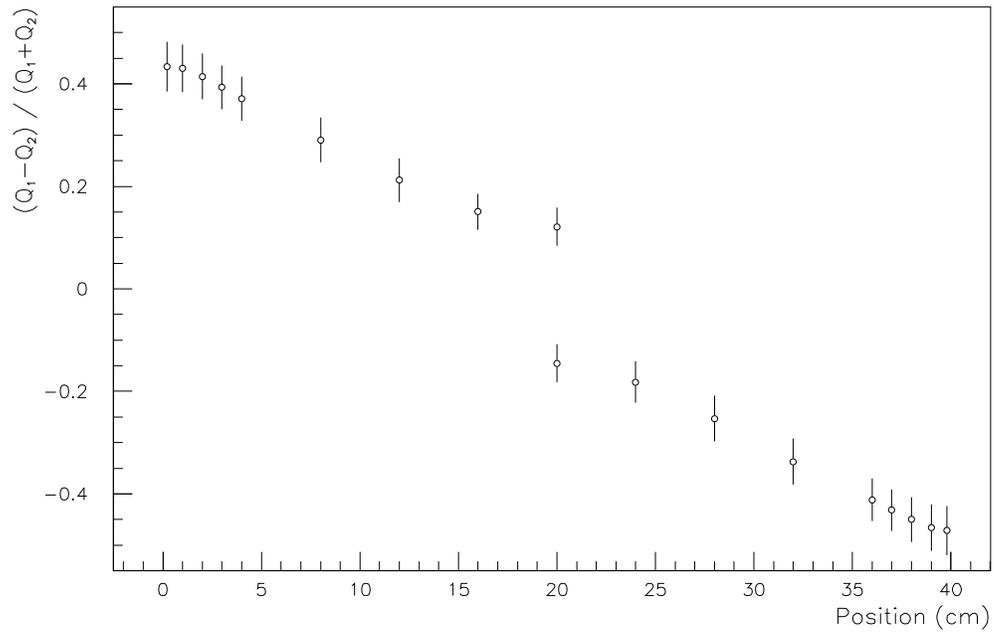,width=15cm}
}
\caption{
The variation of
$\rm { R = (  Q_1 - Q_2 ) / (  Q_1 + Q_2 ) }$
along the longitudinal position of the crystal module,
showing the capability to provide a
position measurement.
}
\label{rvsz}
\end{figure}

The longitudinal position can be obtained
by considering the dimensionless ratio
$\rm { R = (  Q_1 - Q_2 ) / (  Q_1 + Q_2 ) }$,
the variation of which 
at the $^{137}$Cs photo-peak energy
along the crystal length is displayed in Figure~\ref{rvsz}.
The ratio of the RMS errors in R relative to the slope
gives the longitudinal position resolutions.
The measured resolutions 
are 2~cm, 3.5~cm  and 8~cm at 
660~keV, 200~keV and 30~keV, respectively.
The dependence of R on energy is negligible at the
less than the 10$^{-3}$ level. 

\begin{figure}
\centerline{
\epsfig{file=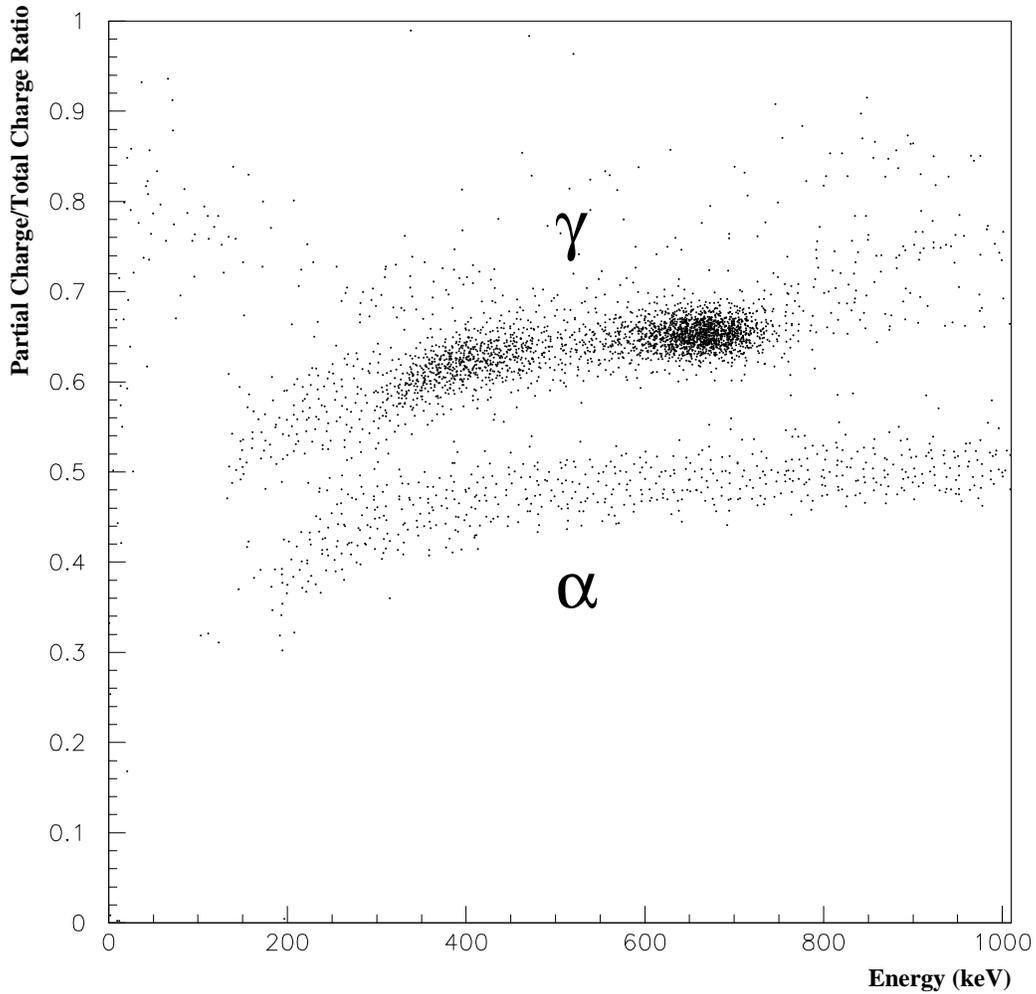,width=15cm}
}
\caption{
The partial charge/total charge ratio in a CsI(Tl)
crystal as a function of energy,
showing excellent pulse shape discrimination
capabilities to differentiate events due to
$\alpha$'s and $\gamma$'s. The $\gamma$-events are
due to a $^{137}$Cs source, showing peaks at the
full-energy and Compton edge regions.
The $\alpha$-events are from the low-energy tails of
an $^{241}$Am-Be source placed on the surface of the
crystal.
}
\label{psdspect}
\end{figure}

In addition, CsI(Tl) provides powerful
pulse shape discrimination (PSD) properties~\cite{csipsd}
to differentiate $\gamma$/e events from those
due to heavily ionizing particles like $\alpha$'s,
which have faster fall time, as shown in Figure~\ref{pulse}.
The typical separation between $\alpha$/$\gamma$ in
CsI(Tl) with the ``Partial Charge Vs Total Charge''
method~\cite{psdmethod} is depicted
in figure~\ref{psdspect}, demonstrating 
an excellent separation of $>$99\% above 500~keV.
Unlike in liquid scintillators, $\alpha$'s are
only slightly quenched in their light output in
CsI(Tl). The quenching factor depends
on the Tl concentration
and the measurement parameters like shaping time:
for full integration of the signals,
the suppression is typically 50\%~\cite{csichar}.

\section{Background Considerations}
\label{sect::bkg}

\subsection{Merits of Crystal Scintillator}
\label{sect::merits}

The suppression, control and understanding
of the background is very important in all 
low background experiments.
The scintillating crystal detector approach
offers several merits~\cite{prospects} to
these ends. The essence are:
\begin{description}
\item[{\bf I.}]
{\bf Large Photon Attenuation:}\\
With its high-Z nuclei, CsI(Tl) provides
very good attenuation to $\gamma$-rays,
especially at the low energy range below
500~keV. 
For instance, the attenuation lengths for a 100~keV
$\gamma$-ray are 0.12~cm and 6.7~cm, respectively,
for CsI(Tl) and liquid scintillator. 
That is, 10~cm of CsI(Tl)
has the same attenuating power as 5.6~m of liquid
scintillator at this low energy.
Consequently,
the effects of external ambient 
$\gamma$ background, like those from
the readout device,
electronics components,
construction materials and
radon diffusion are negligible after
several cm of active veto layer.
Therefore, the background at low energy
will mostly originate within the fiducial
volume due to the internal components.\\
For CsI(Tl) which is non-hygroscopic
and does not need a hermetic seal system to operate, 
{\it ``internal components''} include
only two materials: the crystal itself and the
teflon wrapping sheets, typically at
a mass ratio of 1000:1.
Teflon is known
to have very high radio-purity (typically
better than the ppb level for the
$^{238}$U and $^{232}$Th series)~\cite{hpge}.
As a result,
the experimental challenge becomes
much more focussed - the control and understanding
of the internal radio-purity 
and long-lived cosmic-induced background
of the CsI(Tl) crystal itself.
\item[{\bf II.}]
{\bf Characteristic Detector Response:}\\
The detection threshold is lower while
the energy resolution of CsI(Tl) is better 
than typical
liquid and plastic scintillator with the same modular mass.
Furthermore, in an O(100~kg) CsI(Tl) detector system,
the keV-MeV photons originated within the crystal will
be fully captured. 
These features, together with 
PSD capabilities for $\alpha$-particles
and the granularity of the detector design,
can  provide important diagnostic tools for 
understanding the physical processes 
of the system. Once the background
channels are identified, understood and measured, 
subtraction of its associated effects can be performed.
\end{description}

When 
the dominant background contributions are from
internal contaminations,
two complementary strategies can be
deployed for the background control :
(I) consistent background
subtraction, using the measured spurious
$\alpha$ or $\gamma$ peaks which indicates
residual radioactivity inside the crystal, and
(II) the conventional Reactor ON$-$OFF subtraction.

The background count rate
will be stable and not affected
by external parameters 
such as ambient radon concentrations,
details of the surrounding equipment configurations
and cosmic veto inefficiencies.
Consequently, 
the systematic
uncertainties can be reduced,
and the reliabilities of 
both background suppression processes 
will be more robust. 
In addition,
the large target mass helps to reduce statistical
uncertainties. 
Spectral shape distribution can also be
analyzed to provide additional handles.
For instance the
comparison of the signal rates between
the ``$\rm{< 1 ~ MeV}$''
and the ``$\rm{> 1 ~ MeV}$'' samples
can enhance the sensitivities in the
magnetic moments studies.

\subsection{Background Channels}

The merits discussed above 
allow a compact detector design and
hence, efficient and cost-effective shielding
configurations. While care and the standard
procedures are adopted for suppressing the
ambient radioactivity background (radon purging,
choice of clean construction materials,
photon-counting measurements with germanium detectors,
use of PMT with low-activity glass), the
key background issue remains 
that of {\it internal} background
from the CsI(Tl) itself. The
different contributions 
and their experimental handles
are discussed below.

\subsubsection{Internal Intrinsic Radioactivity}

Unlike in liquid scintillators, $\alpha$'s are
only slightly quenched in their light output in
CsI(Tl).
Crystals contaminated by uranium or thorium would
give rise to multiple peaks above 3~MeV,
as reported in Ref.~\cite{csibkg}.
The absence of multiple peak structures
in our prototype crystals 
suggest a $^{238}$U and $^{232}$Th
concentration of less than
the $10^{-12}$~g/g level [$\sim$1~pkd], 
assuming the decay chains are in equilibrium.
In addition,
direct counting method with a high-purity
germanium detector 
shows the $^{40}$K and $^{137}$Cs 
contaminations of less than the $10^{-10}$~g/g~[$\sim$1700~pkd]
and $4 \times 10^{-18}$~g/g~[$\sim$1200~pkd] levels, respectively.
Mass spectrometry method sets limits of $^{87}$Ru to
less than $8 \times 10^{-9}$~g/g~[$\sim$210~pkd].

Internal radioactivity background typically
consists of $\alpha$, $\beta$ and $\gamma$ emissions
which have 
characteristic energies
and temporal-correlations.
Residual background below the measured
limits can be identified and subtracted off
based on the on-site data.
By careful studies of the
timing and energy correlations among the 
distinct $\alpha$ signatures,
precise information can be obtained on the
radioactive contaminants in the cases where
the $^{238}$U and $^{232}$Th decay series
are not in equilibrium, so that the associated
$\beta$/$\gamma$ background can be accounted for accurately.
For instance, Dark Matter experiments with NaI(Tl)
reported trace contaminations 
(range of $10^{-18} - 10^{-19}$~g/g~[25-250~pkd]
of $^{210}$Pb in the detector, based on $\gamma$-peak at 46.5~keV
and the equivalent peak for $\alpha$'s at
5.4~MeV~\cite{dmnai,wimpi127}. Accordingly,
$\beta$-decays from $^{210}$Bi can be subtracted off
from the signal. 
Similarly, the residual
$\beta$-decays of $^{40}$K and $^{137}$Cs
can be accounted for 
based on their respective characteristic $\gamma$-lines
measure-able from the data.

\subsubsection{Cosmic-Induced Radioactivity}

The experiment is located at a
site with 25~mwe overburden, which is
sufficient
to effectively attenuate the primary
hadronic component from cosmic rays. 
The ``prompt'' cosmic events can
be easily identified, since:
(a) the plastic scintillator veto
can tag them at better than 95\% efficiency,
(b) bremsstrahlung photons from cosmic-rays on the
shieldings cannot reach the inner fiducial
volume of the target, as explained in 
Section~\ref{sect::merits},
(c) cosmic-induced neutrons, mostly from 
the lead, are attenuated and absorbed efficiently
by the boron-loaded polyethylene layers,
and
(d) background
originated from cosmic-rays traversing
the CsI(Tl) target will lead to unmistakably large
pulses ($\sim$20~ MeV for one crystal).

The more problematic background are due to the
long-lived (longer than ms)
unstable isotopes created by the various 
nuclear interaction processes:

\begin{enumerate}
\item {\bf Neutron Capture}

Ambient neutrons or those produced at the 
the lead shieldings have little probability
of being captured by the CsI crystal target,
being attenuated efficiently by the boron-loaded
polyethylene. Cosmic-induced neutrons (energy range MeV) 
originated from the target itself have high probability
of leaving the target.
Residual neutrons can be captured
by the target nuclei $\cs133$
and $\i127$
predominantly via (n,$\gamma$)~\cite{ngtarget}
\begin{eqnarray*}
\rm{n} ~ + ~ ^{133}\rm{Cs} ~ & \rightarrow & ~ ^{134}\rm{Cs} ~~~~
( \rm{\sigma = 30~b ~ ; Q = 6.89~MeV} ) ~ ; \\
\rm{n} ~ + ~ ^{127}\rm{I} ~ & \rightarrow & ~ ^{128}\rm{I} ~~~~
( \rm{\sigma = 6~b ~ ; Q = 6.83~MeV} )
\end{eqnarray*}
with relatively large cross-sections.

The daughter isotope 
$^{134}$Cs ($\rm{\tau_{\frac{1}{2}} = 2.05~yr ~;~ Q = 2.06~MeV }$)
decays with 70\% branching ratio by
beta-decay (end point 658~keV),
plus the emission of two $\gamma$'s (605~keV and 796~keV),
and therefore will not give rise to 
a single hit at the low-energy region.
The isotope 
$^{128}$I ($\rm{ \tau_{\frac{1}{2}} = 25~min~;~ Q = 2.14~MeV }$), 
on the other hand, has
a branching ratio
of 79\% having a lone beta-decay,
which will
mimic the single-hit signature. 
The neutron production rate on-site at
the CsI(Tl) target is estimated to be about
50~pkd. Folding in the capture efficiency
by the target ($\sim 25$\%) and by
$\i127$ in particular ($\sim 14$\%), 
the $^{128}$I production rate is about
1.8~pkd.

The neutron capture rate
by the CsI target can be measured by
tagging $\gamma$-bursts of energy 6.8~MeV.
Knowing the capture rate, the contributions
to the low-energy background due to $^{128}$I
can be evaluated and subtracted off.
Furthermore, the three-fold coincidence from
the $^{134}$Cs decays can be measured, providing
additional information to the neutron capture rates.

\item {\bf Muon Capture}

Cosmic-muons can be stopped by the target nuclei and
subsequently captured~\cite{refmucapture} via 
\begin{eqnarray*}
\mu ^- ~ + ~ \rm{(A,Z)} ~ & \rightarrow & 
~ \rm{(A-Y,Z-1)} ~ + ~ \gamma ~ 's ~
+ ~  \rm{ Y~neutrons } ~~ ,
\end{eqnarray*}
where Y can be 0,1,2,..... with $\rm{ < Y > \sim 1.2 }$.
The daughter isotopes for Y=1,2,3 are all stable, while
the Y=0 case (less than 5\% probability) will give
rise to $^{133}$Xe and $^{127}$Te, both
of which can lead to low-energy single-site background
events: 
$^{133}$Xe 
($\rm{ \tau_{\frac{1}{2}} = 5.3~days~;~ Q = 428~keV }$)
decays with beta (end-point 347~keV) plus
a $\gamma$-ray at 81~keV, while
$^{127}$Te
($\rm{ \tau_{\frac{1}{2}} = 9.4~hr~;~ Q = 690~keV }$)
decays with a lone beta.
The estimated muon capture rate on-site at the CsI target
is $\sim$30~pkd, so that the background
contributions from the Y=0 channels are
less than 1.5~pkd.

\item {\bf Muon-Induced Nuclear Dissociation}

Cosmic-muons can disintegrate the target nuclei
via the ($\gamma$,n) interactions or by 
spallations~\cite{refmudis}, at
an estimated rate of $\sim$10~pkd and $\sim$1~pkd, respectively.
Among the various decay configurations of
the final states nuclei of the ($\gamma$,n) processes,
$^{132}$Cs and $^{126}$I, only about
20\% (or $\sim$2~pkd) of the cases will give rise
to single-hit background. The other decays
give characteristic and identifiable signatures.
For instance, $^{132}$Cs
decays by electron capture resulting in
the emissions of a $\gamma$-ray at 668~keV plus
the X-rays from xenon. 
These can easily
be tagged and used as reference to subtract the 
single-hit background.

\end{enumerate}

\subsubsection{Reactor-ON Correlated Background}

Previous experiments with reactor neutrinos~\cite{refrnuexpt}
as well as on-site measurements
indicate that $\gamma$ and neutron background
associated with ``reactor-ON'' are essentially
zero outside the reactor core and within
reasonable shieldings.
The target region is proton-free
and therefore 
neutrino-induced background from $\nuebar$-p 
is negligible. These interactions, however, will
occur at the polyethylene shieldings.
The prompt e$^+$ will
only give rise at most to 511~keV $\gamma$-rays, 
while the neutron (energy range 1-10~keV)
will be mostly captured by the $^{10}$B in the polyethylene,
producing only 480~keV $\gamma$-rays. Both of these low energy
$\gamma$-background will be efficiently attenuated by
the copper shielding and the active veto.

\subsection{Sensitivity Goals}

From the various
background considerations discussed in
the previous sections, it can be seen that 
while the background control is non-trivial 
like all other low-energy
neutrino experiments, there are more
experimental handles to suppress and identify them
with the crystal scintillator approach.
The efficient $\gamma$-peak detection,
the fine granularity and the PSD capabilities
of the CsI(Tl) detector provides
enhanced analyzing and diagnostic power for the background
understanding. 
The dominating contribution to the
sensitivities is due to the
internal background in the CsI(Tl) target.
The experimental challenge is focussed 
and therefore more elaborate
procedures can be deployed to study and enhance
the radio-purity of 
this one material 
as the experiment evolves.

The present studies place limits on internal
radio-purity to the range of less than the 1000~pkd 
level.
Residual contaminations, if exist, can be
further studied and measured by various methods like
photon counting with germanium detectors, neutron
activation analysis, mass spectroscopy analysis,
as well as by the spectroscopic and time-correlation
input from on-site data taking.
The effects due to cosmic-induced long-lived
isotopes are typically at the range of a few~pkd.
Background due to both channels can be reduced
by consistent background subtraction when the sources
are identified and measured,
and the goal of a suppression
factor of 10$^2$ can be achieve-able.
Such background subtraction strategies
have been successfully used in accelerator
neutrino experiments.
As an illustration, the CHARM-II experiment
measured about 2000 neutrino-electron
scattering events from a
sample of candidate events
with a factor of 20 larger
in size~\cite{charm2}. 
Events due to the 
various background processes were
identified and subtracted off,
such that a few \%
uncertainty in the signal rate had been achieved.

It can be seen
from Figure~\ref{nuerecoil}b that
a Standard Model rate of $\sim$1~pkd can be expected for
a detection threshold of 100~keV.  
After performing the consistent background
suppression,
a residual Background-to-Signal ratio 
of less than 10 before Reactor ON$-$OFF
subtraction is realistic.
Therefore, based on Reactor ON/OFF periods of
100 and 50~days, respectively, 
a fiducial mass of 300~kg of CsI(Tl) target,
and a detector systematic uncertainty of 1\% in
performing the Reactor ON$-$OFF subtraction,
the sensitivity goals
of $\rm{3 \times 10^{-11} ~ \mu_B}$ for the
magnetic moment search and a 5-10\% uncertainty
to the cross-section measurements can be projected.
A comparison with previous on-going experiments
is summarized in 
Table~\ref{tabnueexpt}.

Under similar assumptions, an event rate of
$>$0.005~pkd can be observed
for the $\nuebar$-NCEX channel 
of the various candidate lines 
in the 50-500~keV range,
corresponding to the
cross-section sensitivities
of $\rm{\sim 2 \times 10^{-45} cm^2}$.

\section{Status and Prospects}

By the end of 1999, the design and
prototype studies of the experiment
have been completed. Construction is intensely
underway. A complete 100-kg system,
with full electronics and shieldings,
is expected to be installed on-site
at the Reactor Plant by summer 2000.
Date taking will commence while the
second 100-kg system will be added
in phase. Future upgrades and modifications
of the experiment
will depend on the first results, with
the goals of achieving a 500~kg system
eventually. 

The detector
design adopted in this experiment can
be adopted for other low-energy 
low-background experiments  based
on scintillating crystal detectors~\cite{prospects},
such as Dark Matter WIMP searches~\cite{dmnai,wimpncex}, 
sub-MeV Solar
Neutrino detection (indium-loaded~\cite{snuin}, 
LiI(Eu)~\cite{lii} or GSO~\cite{lensgso} crystals
have been proposed), 
and further studies of
$\nuebar$-NCEX on other isotopes like $^7$Li, $^{10}$B
and $^{11}$B~\cite{b11nuex,nuexaxial}.
Experience and results from the the reactor neutrino
experiment with CsI(Tl) crystals reported
in this work will provide valuable input to these
projects.

Much flexibility is available for detector optimization
based on this generic and easily scale-able 
design.
Different modules can be made of different
crystals.
More and longer crystals can be glued to 
form one module.
Different crystals can be glued together,
in which case the event location
among the various
crystals can be deduced from the different
pulse shape. Passive target can be inserted
to replace a crystal module. New wrapping
materials can be used instead of teflon $-$
there is an interesting new development
with sol-gel coating which can be as thin
as a few microns~\cite{solgel}, thereby even
reducing the passive materials within the
fiducial volume further.

The authors are grateful to the technical staff
of their institutes for the invaluable support,
and to the CYGNUS Collaboration for the loan of the
veto plastic scintillators.
This work was supported by contracts
NSC~87-2112-M-001-034 and NSC~88-2112-M-001-007
from the National Science Council, Taiwan,
as well as  NSF~01-5-23336
and NSF~01-5-23374 from the National Science Foundation, U.S.A.

\clearpage

\input{table1.tab}

\input{table2.tab}

\clearpage

\clearpage

\clearpage

\clearpage

\clearpage

\pagebreak

\clearpage

\clearpage

\clearpage

\end{document}

%% file: table1.tab
\begin{table}
\begin{center}
\begin{tabular}{|l||c|c|c|c|c|}
\hline
& & & & & \\
Properties & CsI(Tl) & NaI(Tl) & BGO & Liquid & Plastic  \\
& & & & &  \\ \hline \hline
& & & & & \\
Density (gcm$^{-3}$) & 4.51 & 3.67 & 7.13 & 0.9 & 1.0  \\
& & & & & \\
Relative Light Yield & 0.45 & 1.00~$^{\dagger}$ & 0.15 & 0.4 & 0.35 \\
& & & & & \\
Radiation Length (cm) & 1.85 & 2.59 & 1.12 & $\sim$45 & $\sim$45 \\
& & & & & \\
dE/dx for MIP (MeVcm$^{-1}$) & 5.6 & 4.8 & 9.2 & 1.8 & 1.9 \\
& & & & & \\
Emission Peak (nm) & 565 & 410 & 480 & 425 & 425 \\
& & & & & \\
Decay Time (ns) & 1000 & 230 & 300 & 2 & 2  \\
& & & & & \\
Refractive index & 1.80 & 1.85 & 2.15 & 1.5 & 1.6  \\
& & & & & \\
Hygroscopic & slightly & yes & no & no & no  \\
& & & & &  \\ \hline
\multicolumn{6}{l}{ }\\
\multicolumn{6}{l}{$^{\dagger}$ Typical light yield for NaI(Tl) 
is about 40000 photons per MeV.}
\end{tabular}
\end{center}
\caption{Characteristic properties of the common
crystal scintillators and their comparison with
typical liquid and plastic scintillators.}
\label{scintab}
\end{table}

%% file: table2.tab
\begin{table}
\small
\begin{tabular}{|l||c|c|c|c|c|}
\hline
Experiment & Savannah\cite{reinesnue} & Kurtchatov\cite{kurtchatovnue}
& Rovno\cite{rovnonue} & NUMU\cite{numunim} & TEXONO-CsI \\ \hline \hline
Target & plastic & fluorocarbon & Si(Li) & CF$_4$ gas & CsI(Tl) \\
& scin. & scin. & & 5 bar & crystals \\
Fiducial mass & 15.9 kg & 103 kg & 37.5 kg & 18.4 kg & 276 kg$^{\ast}$ \\
$\rm{ \phi ( \nuebar ) ~ [ 10^{13}~cm^{-2} s^{-1} ] }$
%%10^{13} ~ cm^{-2} s^{-1} }$ 
& 2.2 & 0.27 & 2.0 & 1.0 & 0.56 \\
Rel. Merit & & & & & \\
\hspace*{0.1cm} (flux $\times$ target) & 
1.0 & 0.7 & 1.8 & 0.44 & 3.1  \\
Threshold (MeV) & 1.5 & 3.1 & 0.6 & 0.5 (goal) & 0.1 (goal) \\
Signal Events / Day & 7.1 & 0.78 & 41 & 10 (expect) &  200 (expect) \\
Event Rates Ratio & & & & & \\
\hspace*{0.1cm} Reactor OFF:(ON$-$OFF) &
5.7 & 9.6 & 120 & 1 (goal) & $<$10 (goal) \\ 
Accuracy of $\rm{ \sigma ( \nuebar e )}$ 
& 29\% & 53\% & 49\% & 5\% (goal)  & 5-10\% (goal)  \\ 
$\mu_{\nu}$ ($\rm{10^{-10}~\mu_B}$)
& 2-4 $^{\dagger}$  & $<$ 2.4 & $<$ 1.9 &  0.3 (goal) 
& 0.3 (goal) \\ \hline
\end{tabular}
\\[2ex]
\small
$^{\ast}$ Based on a 481~kg (16$\times$16 matrix) 
target with 1 veto layer and 5~cm
veto buffer at both ends. \\
$^{\dagger}$ Reanalysis of Savannah River experiment~\cite{vogelengel}
indicates excess of events which can be interpreted as finite
$\mu_{\nu}$.
\vspace*{0.5cm}
\normalsize
\caption{
Characteristics of the previous and current
$\nuebar$-e scattering experiments.
}
\label{tabnueexpt}
\end{table}

%% file: hepex0001.bbl
\clearpage

\begin{thebibliography}{99}
\bibitem{crystal}
For a recent review on the crystal scintillator detector,
see, for example, \\
M. Ishii and M. Kobayashi,
Prog. Crystal Growth and Charact., {\bf 23}, 245 (1991),
and references therein.
\bibitem{emcalo}
For a recent review on the applications of crystal
scintillator in particle physics, see, for example\\
G. Gratta, H. Newman, and R.Y. Zhu,
Ann. Rev. Nucl. Part. Sci. {\bf 44}, 453 (1994).
\bibitem{prospects}
H.T. Wong et al., hep-ex/9910002, submitted to Astropart. Phys. 
\bibitem{pilot}
H..T. Wong and J. Li,
Nucl. Phys. {\bf B} (Proc. Suppl.) {\bf 77}, 177 (1999)
\bibitem{nuint}
For a recent review on experiments on
high energy neutrino
interactions, see, for example,\\
J.M. Conrad, M.H. Shaevitz and T. Bolton,
Rev. Mod. Phys. {\bf 70}, 1341 (1998).
\bibitem{refrnuexpt}
G.~Zacek et al., Phys. Rev. {\bf D 34},
2621 (1986) ;\\
A.I.~Afonin et al., Sov. Phys. JETP {\bf 67} (2), 213 (1988);\\
G.S. Vidyakin et al., JETP Lett. {\bf 59}, 364 (1994); \\
B.~Achkar et al., Nucl. Phys. {\bf B 434}, 503 (1995);\\
Z.D.~Greenwood et al., Phys. Rev. {\bf D 53},
6054 (1996); \\
M.~Apollonio et al., Phys. Lett. {\bf B 420}, 397 (1998);\\
F. Boehm et al., Palo Verde Proposal (1993).
\bibitem{bugeyspect}
B. Achkar et al., Bugey Collaboration, Phys. Lett. {\bf B 374},
243 (1996).

\bibitem{reinesnue}
%%Savahnah River:
F. Reines, H.S. Gurr and H.W. Sobel,
Phys. Rev. Lett. {\bf 37}, 315 (1976).

\bibitem{kurtchatovnue}
%%Kurtchatov : 
G.S. Vidyakin et al, JETP Lett. {\bf 55}, 206 (1992).

\bibitem{rovnonue}
%%Rovno : 
A.I. Derbin et al., JETP Lett. {\bf 57}, 769 (1993).

\bibitem{nudexpt}
T.L. Jenkins, F.E. Kinard, and F. Reines, Phys. Rev.
{\bf 185}, 1599 (1969);
E. Pasierb et al., Phys. Rev. Lett. {\bf 43}, 96 (1979); \\
G.S.~Vidyakin et al., JETP Lett. {\bf 49}, 151 (1988);
G.S.~Vidyakin et al., JETP Lett. {\bf 51}, 279 (1990);\\
S.P. Riley et al., Phys. Rev. {\bf C 59}, 1780 (1999).

\bibitem{bfactories}
Letter of Intent, Belle Coll., KEK (1993);\\
Letter of Intent, BaBar Coll., SLAC-443 (1994).

\bibitem{csichar}
H. Grassman, E. Lorentz, and H.G. Moser,
Nucl. Instrum. Methods {\bf 228}, 323 (1985);\\
P. Schotanus, R. Kamermans, and P. Dorenbos,
IEEE Trans. Nucl. Sci. {\bf 37}, 177 (1990).

\bibitem{sipd}
D.E. Groom, Nucl. Instrum. Methods {\bf 219}, 141 (1984); \\
S. Gunji et al., Nucl. Instrum. Methods {\bf A 295}, 400 (1990); \\
M. Suffert, Nucl. Instrum. Methods {\bf A 322}, 523 (1992).

\bibitem{dmnai}
K. Fushimi et al., Phys. Rev. {\bf C 47}, R425 (1993);\\
M.L. Sarsa et al., Nucl. Phys. {\bf B 35}, 154 (1994);\\
N.J.C. Spooner et al., Phys. Lett. {\bf B 321}, 156 (1994);\\
R. Bernabei et al., Phys. Lett. {\bf B 450}, 448 (1999);\\
G. Gerbier et al., Astropart. Phys. {\bf 11}, 287 (1999).

\bibitem{kyuldjiev}
A.V. Kyuldjiev, Nucl. Phys. {\bf B 243}, 387 (1984).

\bibitem{vogelengel}
P.Vogel and J.Engel, Phys. Rev. {\bf D 39}, 3378 (1989).

\bibitem{kayser}
B. Kayser et al., Phys. Rev. {\bf D 20}, 87 (1979).

\bibitem{nueexptlampf}
R.C. Allen et al., Phys. Rev. Lett. {\bf 55}, 2401 (1985); \\
R.C. Allen et al., Phys. Rev. {\bf D 47}, 11 (1993).

\bibitem{snuexpt}
For a recent review on solar neutrino experiments,
see, for example,\\
R.E. Lanou Jr.,
Nucl. Phys. {\bf B} (Procs. Suppl.) {\bf 77}, 55 (1999).
\bibitem{msw}
L. Wolfenstein, Phys. Rev. {\bf D 17}, 2369 (1978);\\
S.P. Mikheyev and A. Yu. Smirnov, Sov. J. Nucl. Phys. {\bf 42},
1441 (1985).

\bibitem{numunim}
C. Broggini et al., Nucl. Instrum. Methods {\bf A311}, 319 (1992); \\
C. Amsler et al., Nucl. Instrum. Methods {\bf A 396}, 115 (1997).

\bibitem{sn1987nue}
J.M. Lattimer and J. Cooperstein, Phys. Rev. Lett. {\bf 61}, 23 (1988);\\
R. Barbieri and R.N. Mohapatra, Phys. Rev. Lett. {\bf 61} 27 (1988);\\
D. Notzold, Phys. Rev. {\bf D 38}, 1658 (1988).

\bibitem{starcoolnue}
J. Bernstein et al., Phys. Rev. {\bf 132}, 1277 (1963); \\
P. Sutherland et al., Phys. Rev. {\bf D 12}, 2700 (1976); \\
G. Raffelt, Phys. Rev. Lett. {\bf 64}, 2856 (1990).

\bibitem{bbnnue}
J. Morgan, Phys. Lett. {\bf B 102}, 247 (1981);\\
M. Fukugita and S. Yazaki, Phys. Rev. {\bf D 37}, 3817 (1987).

\bibitem{snunue}
M.B. Voloshin, M.I. Vysotskii and L.B. Okun, Sov. Phys. JETP {\bf 64},
446 (1986).

\bibitem{refnueexpt}
I.R. Barabanov et al., Astropart. Phys. {\bf 5}, 159 (1996);\\
A.G. Beda et al., LANL preprint hep-ex/9706004 (1997).

\bibitem{karmennuex}
B. Armbruster et al., Phys. Lett. {\bf B 423}, 15 (1998).

\bibitem{nuex}
H.C. Lee, Nucl. Phys. {\bf A 294}, 473 (1978) ;\\
T.W. Donnelly and R.D. Reccei, Phys. Rep. {\bf 50}, 1 (1979).

\bibitem{b11nuex}
R.S. Raghavan and S. Pakvasa,
Phys. Rev. {\bf D 37}, 849 (1988).

\bibitem{lii}
C.C. Chang, C.Y. Chang, and G. Collins, Nucl. Phys. (Proc. Suppl.)
{\bf B 35}, 464 (1994).

\bibitem{wimpncex}
M.W. Goodman and E. Witten, Phys. Rev. {\bf D 31}, 3059 (1985);\\
J. Ellis, R.A. Flores and J.D. Lewin, Phys. Lett. {\bf B 212}, 375 (1988).

\bibitem{wimpi127}
H. Ejiri, K. Fushimi and H. Ohsumi, Phys. Lett. {\bf B 317}, 14 (1993).

\bibitem{wimpxe129}
P. Belli et al., Phys. Lett. {\bf B 387}, 222 (1996).


\bibitem{nuexaxial}
D.B. Kaplan and A. Manohar, Nucl. Phys. {\bf B 310},527 (1988);\\
J. Bernab\'{e}u  et al., Nucl. Phys. {\bf B 378}, 131 (1992);\\
G. Garvey et al., Phys. Rev. {\bf C 48}, 1919 (1993); \\
K. Kubodera and S. Nozawa, Int. J. Mod. Phys. {\bf E 3}, 101 (1994).

\bibitem{orland}
ORLaND Proposal (1998).

\bibitem{earthnu}
L.M. Krauss, S.L. Glashow and D.N. Schramm, Nature {\bf 310}, 191 (1984).

\bibitem{taiwan} 
C.Y. Chang, S.C. Lee and H.T. Wong,
Nucl. Phys. {\bf B} (Procs. Suppl.) {\bf 66}, 419 (1998).

\bibitem{csipsd}
J. Alarja et al., Nucl. Instrum. Methods {\bf A 242}, 352 (1986); \\
P. Kreutz et al., Nucl. Instrum. Methods {\bf A 260}, 120 (1987).

\bibitem{psdmethod}
C.L. Morris et. al., Nucl. Instrum. Methods {\bf 137},
397 (1976);\\
M.S. Zucker and N. Tsoupas,  Nucl. Instrum. Methods {\bf A299},
281 (1990).

\bibitem{hpge}
P. Jagam and J.J. Simpson, Nucl. Instrum. Methods {\bf A 324},

\bibitem{csibkg}
U. Kilgus, R. Kotthaus, and E. Lange, Nucl. Instrum. Methods
{\bf A 297}, 425, (1990); \\
R. Kotthaus, Nucl. Instrum. Methods {\bf A 329}, 433 (1993).


\bibitem{ngtarget}
V.L. Alexeev et al., Nucl. Phys. {\bf A248}, 249 (1975);\\
L.A. Schaller, J. Kern and B. Michaud, Nucl. Phys. {\bf A 165}, 415 (1971).

\bibitem{refmucapture}
S. Charalambus, Nucl. Phys. {\bf A 166}, 145 (1971);\\
T. Suzuki, D.F. Measday, and J.P. Roalsvig,
Phys. Rev. {\bf C 35}, 2212 (1987);\\
T. Kozlowski et al., Nucl. Phys. {\bf A 436}, 717 (1985).

\bibitem{refmudis}
G. Cocconi and V. Cocconi Tongiorgi, Phys. Rev {\bf 84}, 29 (1951);\\
S. Hayakawa, Phys. Rev {\bf 84}, 37 (1951).

\bibitem{snuin}
R.S. Raghavan, Phys. Rev. Lett. {\bf 37}, 259 (1976);\\
M. Avenier et al., Nucl. Phys. $\bf B$ (Proc. Suppl.)
{\bf 28A}, 496 (1992).

\bibitem{lensgso}
R.S. Raghavan, Phys. Rev. Lett. {\bf 78}, 3618 (1997).

\bibitem{solgel}
R. Chipaux et al., DAPNIA-SED-98-01,
to be published in IEEE Trans Nucl. Sci. (1999).

\bibitem{charm2}
P. Vilain et al., Phys. Lett. {\bf B 335}, 246 (1994).

\end{thebibliography}
